# Wave function forms of interlayer excitons in bilayer transition metal dichalcogenides

Jianju Tang[1†], Songlei Wang[2†], Yuhang Hou[3], Hongyi Yu[3,4*]

[1] School of Physics and Electronic Engineering, Hanshan Normal University, Chaozhou 521000, China
[2] School of Physics, Peking University, Beijing 100871, China
[3] Guangdong Provincial Key Laboratory of Quantum Metrology and Sensing & School of Physics and Astronomy, Sun Yat-Sen University (Zhuhai Campus), Zhuhai 519082, China
[4] State Key Laboratory of Optoelectronic Materials and Technologies, Sun Yat-Sen University (Guangzhou Campus), Guangzhou 510275, China

[†] These authors contribute equally to this work
[*] E-mail: yuhy33@mail.sysu.edu.cn

**Abstract:** We numerically solve the electron-hole relative wave function of interlayer excitons in bilayer transition metal dichalcogenides, taking into account the screening effects from both the constituent transition metal dichalcogenides layers and the surrounding dielectric environment. We find that the wave function of the 1s ground state is close to the gaussian form, rather than the well-known exponential decay form of the two-dimensional hydrogen model. Meanwhile, the 2s state has an energy $E_{2s}$ significantly higher than $E_{2p}$ of the 2p state, but becomes close to $E_{3d}$ of the 3d state with $E_{2s} - E_{2p} \approx E_{3d} - E_{2p} \approx E_{2p} - E_{1s}$ under a large interlayer separation and weak environmental screening. Under general conditions, the solved 1s, 2p and 3d wave functions can be fit nearly perfectly by simple analytic forms which smoothly cross from gaussian to exponential decay. These analytic forms can facilitate the accurate evaluation of various exciton quantities for device applications.

In the past decade, monolayer transition metal dichalcogenides (TMDs) with direct band gaps in the visible frequency range have emerged as a promising two-dimensional (2D) platform for exploring next-generation optoelectronic applications. Due to the reduced screening of the layered geometry and the large effective mass of band-edge carriers, the optical properties of monolayer TMDs are dominated by excitons [1-3], which are electron-hole pairs bound by the strong Coulomb interaction. Excitons in monolayer TMDs are found to exhibit large binding energies in the order of several hundred meV [4,5] and Bohr radii as small as 1 to 2 nm [6-10]. Besides, the dielectric constant of the surrounding environment is found to greatly affect the exciton binding energy and Bohr radius [11-18], which can be used to engineer the exciton properties. Stacking two TMDs monolayers to form van der Waals bilayer structures opens up a new realm to extend their already extraordinary properties. Interlayer excitons (IXs) with the electron and hole constituents located in different TMDs layers have been detected in various heterobilayer and homobilayer TMDs structures [19-21], which host great tunability due to the finite out-of-plane electric dipoles. Numerical calculations and experimental measurements have been carried out to obtain the binding energies of IXs in various bilayer TMDs systems, which are found to be in the order of 100 meV [21-27]. The corresponding Bohr radii around 2 nm have also been measured experimentally [10,28]. Similar to the 2D hydrogen model, the electron-hole relative motion of these excitons forms a series of discrete Rydberg states. It has been shown that, due to the nonlocal dielectric screening of the atomically-thin layered structure [29], the energies of Rydberg states in monolayer TMDs deviate significantly from $E_{nm}^{(2DH)} \propto -(n-1/2)^{-2}$ of the well-known 2D hydrogen model. Here $n = 1, 2, \ldots$ is the principal quantum number and $m = 0, \pm 1, \pm 2, \ldots$ is the angular momentum. Various analytical formulations have been adopted to approximate the Rydberg state energies in monolayer TMDs using different approaches [14,30,31]. However, for IXs in bilayer TMDs, such analytical formulas are still lacking.

Besides the energy and Bohr radius, other important IX properties, including the oscillator strength [32,33], external fields induced coupling between Rydberg orbitals [27,34,35], geometric structures [36], etc., can play important roles in exciton-based optoelectronic and valleytronic applications. For accurate evaluations of these quantities, it is essential to know the wave function form of the electron-hole relative motion. Some previous works used exponential decay forms similar to those in the 2D hydrogen model [37-39]. However, it is known that the nonlocal dielectric screening of the 2D geometry leads to a significant deviation of the exciton in layered TMDs to the 2D hydrogen model [29]. Combined with the finite interlayer separation, IX wave functions can be significantly different from the exponential decay forms of the 2D hydrogen model. Meanwhile, some other literatures have treated the interlayer Coulomb potential as a harmonic potential [40,41], where low-energy IXs can be described by 2D harmonic oscillators. Such a harmonic approximation, however, usually requires the interlayer separation $D$ to be much larger than the exciton Bohr radius $a_B$, which is not satisfied in most bilayer TMDs systems where $D \sim 6$ Å and $a_B \sim 2$ nm [8-10,28,42]. Thus, further analysis is needed for more accurate descriptions of IX wave functions in bilayer TMDs, especially analytically tractable forms which are easy to implement for device applications.

In this work, we focus on the wave functions of four lowest-energy Rydberg states (1s, 2p, 2s and 3d) for IXs in bilayer TMDs. Using a Coulomb potential form modified by the dielectric screening of the two TMDs layers as well as the encapsulating thick hexagonal boron nitride (hBN) layers, we numerically solve the electron-hole relative wave function of the IX. We find that the 1s ground state can be approximated by a harmonic oscillator with the wave function close to a gaussian form. Under a large interlayer separation $D$ and weak environmental dielectric screening effect, 2s state has an energy significantly higher than that of 2p state, but close to the energy of 3d state with $E_{2s} \approx E_{3d}$ and $E_{2s} - E_{2p} \approx E_{2p} - E_{1s}$. These come from the combined effect of the finite interlayer separation and nonlocal screening of layered TMDs. Under small $D$ values and with the presence of hBN encapsulation, neither gaussian nor exponential decay can well describe the exciton wave function. Nevertheless, we show that there exist simple analytic forms which smoothly cross from gaussian to exponential decay and always fit nearly perfectly to the numerically solved 1s, 2p and 3d wave functions.

We consider a TMDs bilayer separated by a vertical distance $D$, which is $\approx 6$ Å and can be further increased by inserting a few hBN layers. Meanwhile, the bilayer can be encapsulated by a thick substrate and a capping hBN layers, whose dielectric constants can both be approximated as $\epsilon \approx 4.5$ (the average dielectric constant of bulk hBN). The interface of the substrate (capping layer) is vertically separated by a distance $d'$ from the lower (upper) TMDs layer, see Fig. 1(a) for an illustration. An electron and a hole located in different TMDs layers interact through the interlayer Coulomb potential $V(r)$, with $r = |\mathbf{r}|$ the electron-hole in-plane distance. Modified by the atomically-thin geometry of TMDs layers as well as dielectric screenings from the substrate and capping layers, $V(r)$ takes the following form (see Appendix I for a detailed derivation)

$$V(r) = \int_0^\infty dq \frac{-J_0(qr)e^{-qD}MQ}{[Q + (N - e^{-qD}M)qr_0][Q + (N + e^{-qD}M)qr_0]}. \tag{1}$$

Here $J_0$ is the Bessel functions of the first kind, $M = \left(1 - \frac{\epsilon-1}{\epsilon+1}e^{-2qd'}\right)^2$, $Q = \left(1 - \frac{\epsilon-1}{\epsilon+1}e^{-q(D+2d')}\right)\left(1 + \frac{\epsilon-1}{\epsilon+1}e^{-q(D+2d')}\right)$, $N = \left(1 - \frac{\epsilon-1}{\epsilon+1}e^{-2q(D+d')}\right)\left(1 - \frac{\epsilon-1}{\epsilon+1}e^{-2qd'}\right)$. $r_0 \approx 4.5$ nm corresponds to the screening length of monolayer TMDs. Obviously, $V(r)|_{r_0 \to 0, d' \to \infty} = \frac{-1}{\sqrt{r^2+D^2}}$ becomes the traditional form in a homogeneous 3D space. Meanwhile for vanishing interlayer separations, $V(r)|_{D \to 0, d' \to 0} = \frac{-1}{\epsilon}\int_0^\infty dq \frac{J_0(qr)}{1+2qr_0/\epsilon} = \frac{-\pi}{4\epsilon r_0}\left[H_0\left(\frac{\epsilon r}{2r_0}\right) - Y_0\left(\frac{\epsilon r}{2r_0}\right)\right]$ becomes the monolayer Keldysh form with a screening length $2r_0/\epsilon$. Note that in early works [22], an interlayer Coulomb potential form $W(r) = \frac{-1}{\epsilon}\int_0^\infty dq \frac{e^{-qD}J_0(qr)}{[1+(1+e^{-qD})qr_0/\epsilon][1+(1-e^{-qD})qr_0/\epsilon]}$ different from $V(r)$ has been used (see Fig. 2(a) for a comparison between $V(r)$ and $W(r)$). The form of $W(r)$ is applicable when the entire 3D space is covered by the dielectric medium $\epsilon$, whereas Eq. (1) we used has considered the fact that the dielectric medium cannot extend to the region $z \in [-D/2 - d', D/2 + d']$ where the TMDs bilayer resides (see Fig. 1(a)). Below we consider two cases: (1) $d' \to \infty$, which corresponds to TMDs suspended in vacuum; (2) $d' \approx 5$ Å, which corresponds to TMDs encapsulated by thick hBN layers.

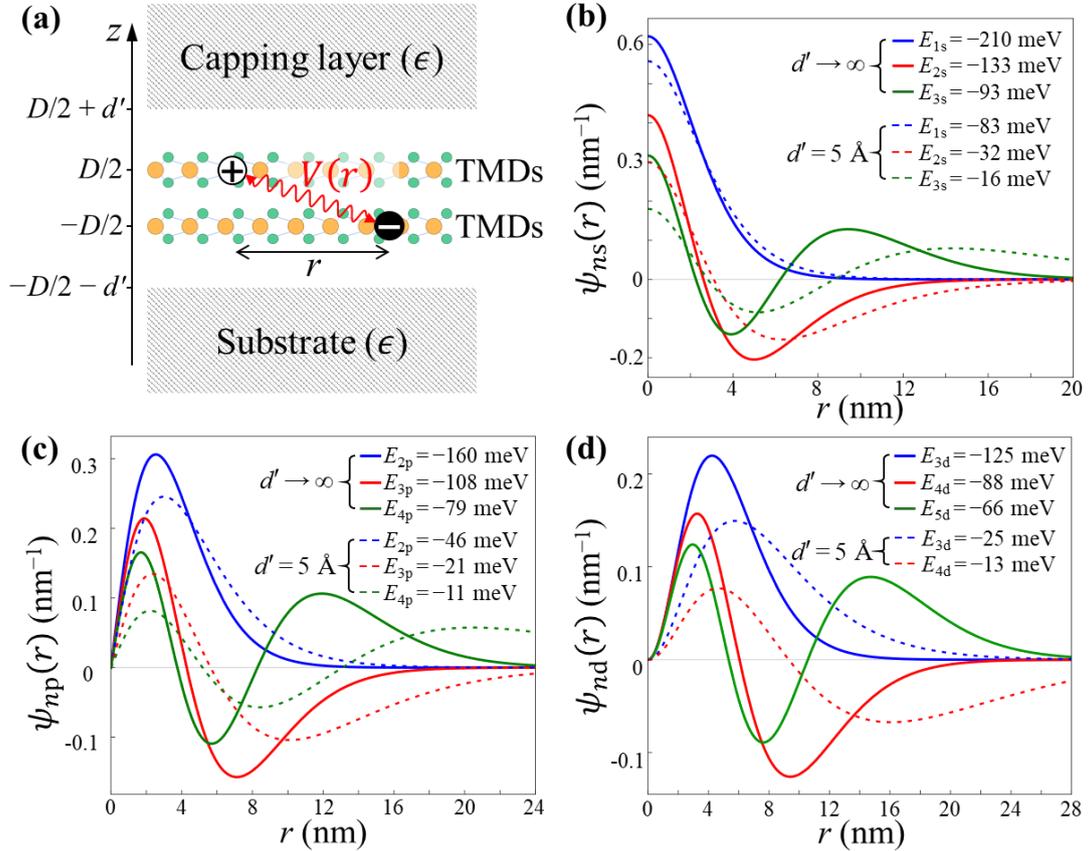

**Fig. 1** (a) A schematic illustration of the bilayer TMDs system encapsulated by a capping layer and a substrate with dielectric constants $\epsilon$. The two TMDs monolayers are separated by a vertical distance $D$, and the top (bottom) interface of the substrate (capping layer) is at a distance $d'$ away from the lower (upper) TMDs layer. (b-d) Numerically solved radial wave functions of several lowest-energy (b) s-type exciton states with $m = 0$, (c) p-type states with $m = \pm 1$, (d) d-type states with $m = \pm 2$. Both the suspended case with $d' \to \infty$ and hBN-encapsulated case with $\epsilon \approx 4.5$, $d' \approx 5$ Å are shown. Other parameters are set as $D = 6$ Å and $r_0 = 4.5$ nm.

Considering the rotational symmetry of $V(r)$, the IX eigenstate with an energy $E_{nm}$ is in the form $\Psi_{nm}(\mathbf{r}) \equiv \frac{e^{im\theta}}{\sqrt{2\pi}} \psi_{nm}(r)$, where the angular quantum number $m = 0, \pm 1, \pm 2, \ldots$ correspond to eigenstates of s-type, p-type, d-type, …, respectively. Following the notation of the 2D hydrogen model, the lowest-energy states of $m = 0, \pm 1, \pm 2$ are denoted as 1s, 2p, 3d states, respectively. The radial wave function $\psi_{nm}(r)$ satisfies the Schrödinger equation

$$\left[-\frac{\hbar^2}{2\mu}\left(\frac{\partial^2}{\partial r^2} + \frac{1}{r}\frac{\partial}{\partial r} - \frac{m^2}{r^2}\right) + V(r)\right]\psi_{nm}(r) = E_{nm}\psi_{nm}(r). \quad (2)$$

Here $\mu \approx 0.25 m_0$ is the reduced mass of the exciton with $m_0$ the free electron mass. Obviously, $E_{nm}$ and $\psi_{nm}$ are independent on the sign of $m$. We show the numerically solved wave functions for several lowest-energy eigenstates of $m = 0, \pm 1$ and $\pm 2$ in Fig. 1(b), 1(c) and 1(d), respectively, under typical parameters of $D = 6$ Å and $r_0 = 4.5$ nm. The corresponding energies are also given. Under $d' \to \infty$, the energy separation $E_{2p} - E_{1s} \approx 50$ meV between the ground state 1s and first excited state 2p agrees well with the experimental measured value of 67 meV [27]. Such an energy separation is only a small fraction of the exciton binding energy $E_b \approx 210$ meV, in sharp contrast to the well-known 2D hydrogen model with $E_{2p} - E_{1s} = \frac{8}{9} E_b$.

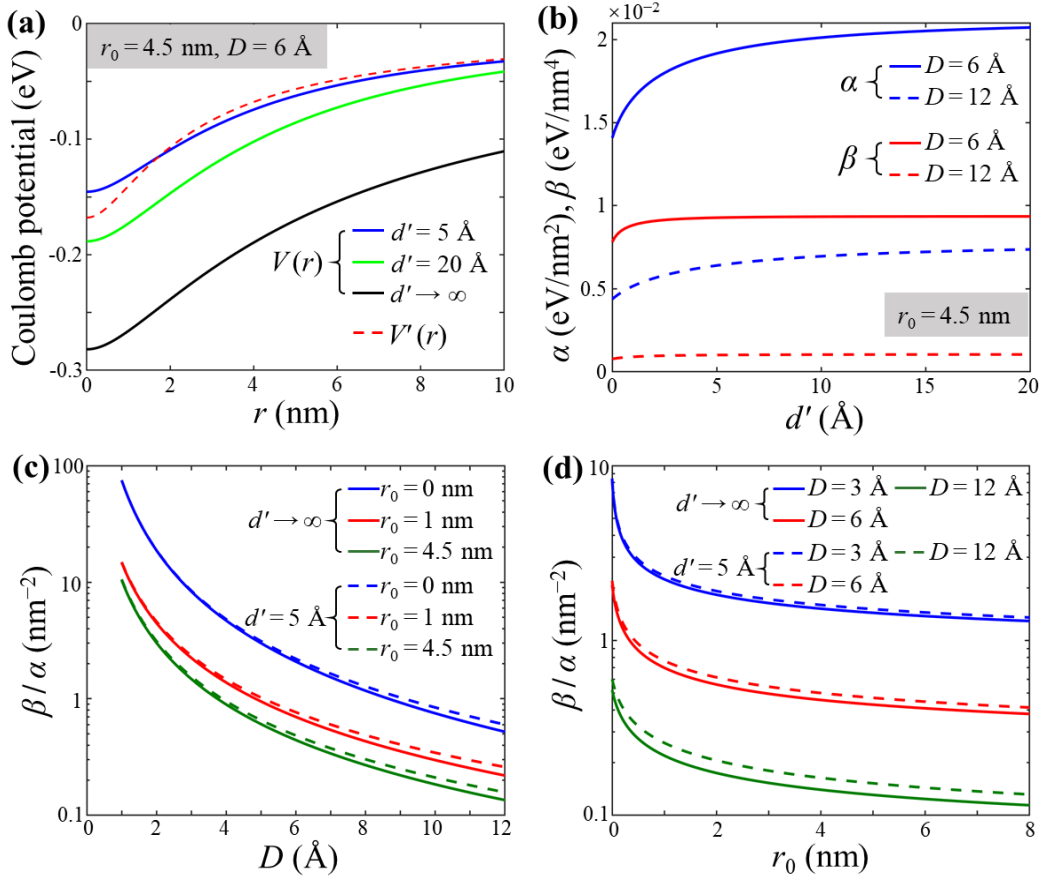

**Fig. 2** (a) The interlayer Coulomb potential as a function of the electron-hole in-plane separation $r$ under $r_0 = 4.5$ nm and $D = 6$ Å. The solid curves correspond to $V(r)$ from Eq. (1) under three different values of $d'$. The red dashed curve is another Coulomb potential form $W(r)$, which is applicable when the entire 3D space is covered by the dielectric medium with a dielectric constant $\epsilon$. (b) The harmonic confinement strength $\alpha$ and the strength $\beta$ of the lowest-order anharmonicity (see Eq. (3) in the maintext) as functions of $d'$, under $r_0 = 4.5$ nm and $D = 6, 12$ Å. (c) The ratio $\beta/\alpha$ as a function of $D$ under $r_0 = 0, 1, 4.5$ nm. (d) $\beta/\alpha$ as a function of $r_0$ under $D = 3, 6$ and 12 Å.

As shown in Fig. 1(b), the wave function $\psi_{1s}$ barely changes when increasing $d'$ from 5 Å to $\infty$, implying that it is insensitive to the environmental dielectric screening. However, the corresponding energy $E_{1s}$ changes significantly from $-83$ to $-210$ meV. To gain insights about this contradiction, we show curves of $V(r)$ under several different values of $d'$ in Fig. 2(a). For $r < 2$ nm which corresponds to the typical electron-hole in-plane separation of 1s state in monolayer/bilayer TMDs [8-10,28,42], $V(r)$ is analogous to a harmonic potential due to the finite interlayer separation $D$. We then expand $V(r)$ near $r = 0$ as

$$V(r) = V(0) + \alpha r^2 - \beta r^4 + O(r^6). \tag{3}$$

$\alpha$ is the harmonic confinement strength, and $\beta$ corresponds to the strength of the lowest-order anharmonicity. We can see from Fig. 2(a) that the lowest potential value $V(0)$ raises fast with the decrease of $d'$, which is the origin of the large binding energy difference between the suspended ($d' \to \infty$) and hBN-encapsulated bilayer TMDs ($d' = 5$ Å). On the other hand, Fig. 2(b) indicates that both $\alpha$ and $\beta$ are insensitive to $d'$ when $d' \geq 5$ Å, which then explains why the wave function $\psi_{1s}$ barely changes when $d'$ is increased from 5 Å to $\infty$.

We emphasize that Eq. (3) and the related conclusions only applies to IX states with narrow spatial extensions such that $\alpha/\beta \gg \langle r^2 \rangle$, which does not apply to higher-energy excited states with large spatial extensions. IX states satisfying this condition can be well approximated by a 2D harmonic oscillator model, with a wave function in the gaussian form rather than the exponential decay form of the well-known 2D hydrogen model. This requires the value $\beta/\alpha$ to be small. Fig. 2(c) and 2(d) show our calculated $\beta/\alpha$ as a function of $D$ and $r_0$, respectively. $\beta/\alpha$ is found to decreases sharply with the increase of $D$, and also drops sharply when increasing $r_0$ from 0 to several nm but becomes slowly decreasing with the further increase of $r_0$. However, increasing $D$ and $r_0$ also weakens the Coulomb interaction thus can result in a larger spatial extension of the electron-hole relative motion, which leads to a more important role for the anharmonic term $\beta r^4$. The competition between these two mechanisms then determines whether the harmonic oscillator model is a good approximation.

Below we focus on the four lowest-energy IX states, namely 1s, 2s, 2p and 3d. Fig. 3(a) summarizes our calculated energies of these states as functions of interlayer distance $D$ for TMDs bilayers suspended in vacuum with $d' \to \infty$. With the increase of $D$, the energies of 2s and 3d states become more and more close, and the splittings between 1s, 2p and 3d satisfy $E_{2p} - E_{1s} \approx E_{3d} - E_{2p}$ at large $D$ values (see the dashed line in Fig. 3(a), which corresponds to the value $2E_{2p} - E_{1s}$). Such a behavior is distinct from the 2D hydrogen model with $E_{2s} = E_{2p}$, but consistent with the 2D harmonic oscillator model where $E_{2p} - E_{1s} = E_{3d} - E_{2p} = E_{2s} - E_{2p}$. This finding is further corroborated by our calculated root-mean-square electron-hole separation $\sqrt{\langle r^2 \rangle}$ shown in Fig. 3(b). $\sqrt{\langle r^2 \rangle}$ increases slowly with $D$ and is around 2 nm for the 1s ground state, in qualitative agreement with experimentally measured values [8-10,28,42]. For hBN-encapsulated bilayer TMDs with $d' = 5$ Å, we show $E_{nm}$ and $\sqrt{\langle r^2 \rangle}$ in Fig. 3(c) and 3(d), respectively. Compared to the suspended case in Fig. 3(a), the value $2E_{2p} - E_{1s}$ shown as dashed line in Fig. 3(c) has a larger deviation to $E_{3d}$ and $E_{2s}$. Meanwhile in Fig. 3(d), $\sqrt{\langle r^2 \rangle}$ of 2s, 2p and 3d excited states are significantly larger than those in Fig. 3(b), but $\sqrt{\langle r^2 \rangle}$ of 1s is nearly the same. This is consistent with experimental observations that the absorption of 1s (2s) exciton is insensitive (rather sensitive) to the environmental dielectric screening [12,43]. Fig. 3(e) and 3(f) show how $E_{nm}$ and $\sqrt{\langle r^2 \rangle}$, respectively, change with $r_0$. Since the 2D dielectric screening of layered TMDs weakens the Coulomb interaction, $|E_{nm}|$ decreases and $\sqrt{\langle r^2 \rangle}$ becomes larger with the increase of $r_0$. On the hand, the agreement between $2E_{2p} - E_{1s}$ and $E_{3d}, E_{2s}$ gets better for larger $r_0$, see the dashed line in Fig. 3(e).

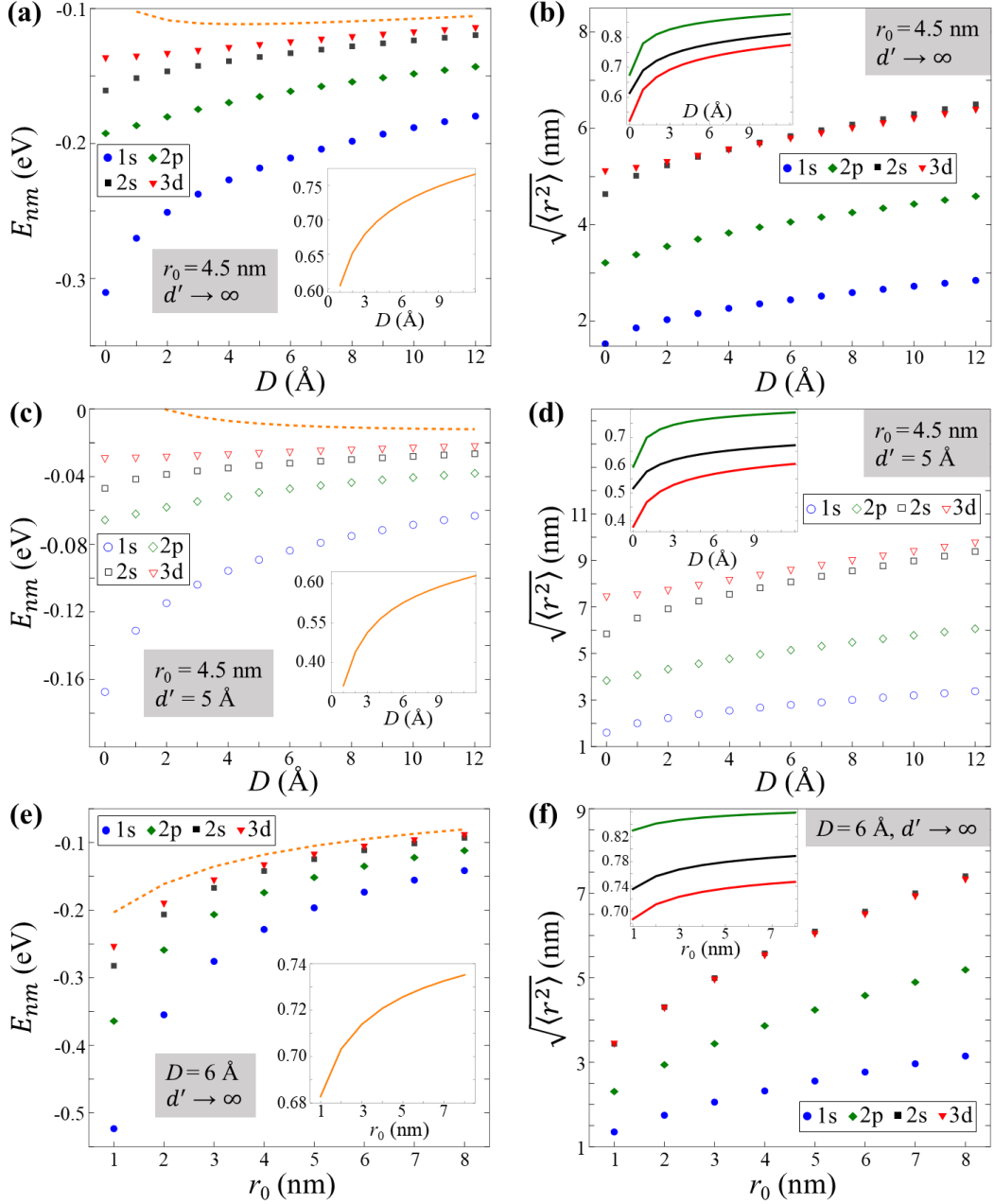

**Fig. 3** The calculated (a) energies and (b) root-mean-square electron-hole separations $\sqrt{\langle r^2 \rangle}$ of 1s, 2s, 2p and 3d states as functions of $D$, for a suspended TMDs bilayer with $d' \to \infty, r_0 = 4.5$ nm. (c) and (d) show $E_{nm}$ and $\sqrt{\langle r^2 \rangle}$, respectively, but for hBN-encapsulated TMDs bilayers with $d' = 5$ Å and $r_0 = 4.5$ nm. (e) and (f) are $E_{nm}$ and $\sqrt{\langle r^2 \rangle}$, respectively, as functions of $r_0$ under $D = 6$ Å and $d' \to \infty$. In (a), (c) and (e), the dashed orange lines show the energy positions $2E_{2p} - E_{1s}$, and the solid orange lines in the insets correspond to the ratio $(E_{3d} - E_{2p})/(E_{2p} - E_{1s})$. In the insets of (b), (d) and (f), the solid green, black and red lines correspond to $\sqrt{2\langle r^2 \rangle_{1s}}/\sqrt{\langle r^2 \rangle}_{2p}$, $\sqrt{\frac{17}{5}\langle r^2 \rangle_{1s}}/\sqrt{\langle r^2 \rangle}_{2s}$ and $\sqrt{3\langle r^2 \rangle_{1s}}/\sqrt{\langle r^2 \rangle}_{3d}$, respectively, which all equal 1 in the 2D harmonic oscillator model.

Note that $E_{2p} - E_{1s} = E_{3d} - E_{2p}$ and $\sqrt{\langle r^2 \rangle}_{1s} : \sqrt{\langle r^2 \rangle}_{2p} : \sqrt{\langle r^2 \rangle}_{3d} : \sqrt{\langle r^2 \rangle}_{2s} = 1 : \sqrt{2} : \sqrt{3} : \sqrt{17/5}$ in the 2D harmonic oscillator model. We then use the ratios $(E_{3d} - E_{2p})/(E_{2p} - E_{1s})$, $\sqrt{2\langle r^2 \rangle_{1s}}/\sqrt{\langle r^2 \rangle}_{2p}$, $\sqrt{\frac{17}{5}\langle r^2 \rangle_{1s}}/\sqrt{\langle r^2 \rangle}_{2s}$ and $\sqrt{3\langle r^2 \rangle_{1s}}/\sqrt{\langle r^2 \rangle}_{3d}$ as qualitative measures to the agreements between the exciton states and harmonic oscillator model. The results of $(E_{3d} - E_{2p})/(E_{2p} - E_{1s})$ are shown as solid orange lines in the insets of Fig. 3(a,c,e), whereas $\sqrt{2\langle r^2 \rangle_{1s}}/\sqrt{\langle r^2 \rangle}_{2p}$, $\sqrt{\frac{17}{5}\langle r^2 \rangle_{1s}}/\sqrt{\langle r^2 \rangle}_{2s}$ and $\sqrt{3\langle r^2 \rangle_{1s}}/\sqrt{\langle r^2 \rangle}_{3d}$ are shown as solid green, black and red lines, respectively, in the insets of Fig. 3(b,d,f). These ratios are all found to approach unity with the increase of $D$ or $r_0$, especially for the $d' \to \infty$ case. We thus expect the 1s wave function to be close to a gaussian form under large $D$ or $r_0$ values. On the other hand, the excited states have larger spatial extensions, which are then expected to have poorer agreement with the harmonic oscillator model than the 1s state. For $d' = 5$ Å, the agreement for the ratio between $E_{nm}$ or $\sqrt{\langle r^2 \rangle}$ with that of the 2D harmonic oscillator model is not as good as in Fig. 3(a,b), but still deviates strongly from the results of the 2D hydrogen model where $E_{2s} = E_{2p} = \frac{1}{9}E_{1s}$ and $\sqrt{\langle r^2 \rangle}_{1s} : \sqrt{\langle r^2 \rangle}_{2p} : \sqrt{\langle r^2 \rangle}_{3d} : \sqrt{\langle r^2 \rangle}_{2s} = 1 : \sqrt{30} : \sqrt{175} : \sqrt{39}$ (see Fig. 3(c,d)).

For a more quantitative investigation, we use three analytic equation forms to approximate the 1s wave function: a gaussian form $\psi_{1s}^{(g)}(r)$, an exponential decay form $\psi_{1s}^{(e)}(r)$ and a transitional form $\psi_{1s}^{(t)}(r)$. They are given by

$$\psi_{1s}^{(g)}(r) = \frac{2}{A}e^{-r^2/A^2},$$
$$\psi_{1s}^{(e)}(r) = \frac{2}{B}e^{-r/B}, \quad (4)$$
$$\psi_{1s}^{(t)}(r) = \frac{2}{\sqrt{C'^2 + 2CC'}}e^{-(\sqrt{r^2+C^2}-C)/C'}.$$

Here $\psi_{1s}^{(t)}(r)\big|_{r \ll C} \to \frac{2}{\sqrt{C'^2+2CC'}}e^{-\frac{r^2}{2CC'}}$ and $\psi_{1s}^{(t)}(r)\big|_{r \gg C} \to \frac{2}{\sqrt{C'^2+2CC'}}e^{-\frac{r}{C'}}$, thus is termed as the transitional form. $\psi_{1s}^{(t)}$ has been found to be a reasonably accurate variational function for the ground state exciton in coupled quantum well structures [44]. By fitting to the numerically calculated $r|\psi_{1s}(r)|^2$, the parameters $A$, $B$, $C$ and $C'$ can be obtained. The corresponding fitting curves of $\psi_{1s}^{(g)}$, $\psi_{1s}^{(e)}$ and $\psi_{1s}^{(t)}$ are indicated in Fig. 4 as solid lines, along with the numerical data of $\psi_{1s}$ and the coefficients of determination $R^2$. We can see that the exponential form $\psi_{1s}^{(e)}$ deviates significantly from $\psi_{1s}$, whereas the gaussian form $\psi_{1s}^{(g)}$ can serve as a good approximation to $\psi_{1s}$ for $D = 6$ Å or larger. The agreement between $\psi_{1s}^{(g)}$ and $\psi_{1s}$ gets better with the increase of $D$ (Fig. 4(a,b)) or $r_0$ (Fig. 4(c)), but becomes slightly worse when $d'$ is decreased from $\infty$ to 5 Å (Fig. 4(b)). Meanwhile, the transitional form $\psi_{1s}^{(t)}$ always gives nearly perfect fittings for all cases under various sets of parameters. These observations suggest that compared to the exponential decay variational wave function, a gaussian one can give more accurate results for ground state properties of IXs in TMDs where $D \gtrsim 6$ Å. If one further wishes to achieve the best accuracy, a transitional form $\psi_{1s}^{(t)}$ given in Eq. (4) with two variational parameters can be used.

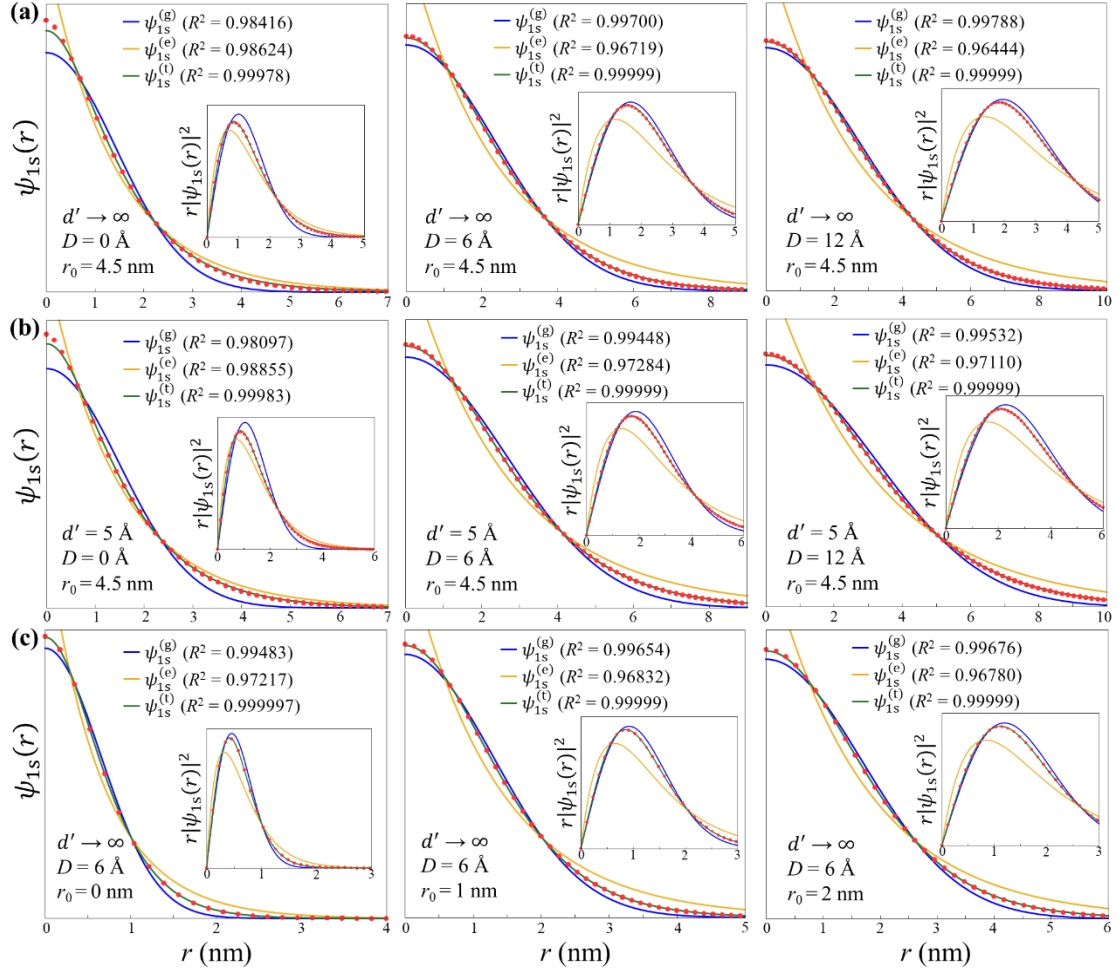

**Fig. 4** The numerically calculated 1s wave functions $\psi_{1s}$ (red dots) and the approximated gaussian form $\psi_{1s}^{(g)}$ (blue lines), exponential decay form $\psi_{1s}^{(e)}$ (yellow lines) and transitional form $\psi_{1s}^{(t)}$ (green lines). The insets show the fitting results to $r|\psi_{1s}(r)|^2$ (red dots) using $r\left|\psi_{1s}^{(g)}(r)\right|^2$ (blue lines), $r\left|\psi_{1s}^{(e)}(r)\right|^2$ (yellow lines) and $r\left|\psi_{1s}^{(t)}(r)\right|^2$ (green lines). (a) is under $r_0 = 4.5$ nm, $d' \to \infty$ and three different values of $D$; (b) is under $r_0 = 4.5$ nm, $d' = 5$ Å and three different values of $D$; (c) is under $d' \to \infty$, $D = 6$ Å and three different values of $r_0$.

The transitional equation form can be further generalized to 2p and 3d excited states, with:

$$\psi_{2p}^{(t)}(r) = \frac{4/C'}{\sqrt{6C'^2+12CC'+8C^2}} re^{-(\sqrt{r^2+C^2}-C)/C'},$$
$$\psi_{3d}^{(t)}(r) = \frac{2\sqrt{2}/C'}{\sqrt{15C'^4+30CC'^3+24C^2C'^2+8C^3C'}} r^2 e^{-(\sqrt{r^2+C^2}-C)/C'}. \tag{5}$$

We then use $\psi_{2p}^{(g)}(r) = \frac{2\sqrt{2}}{A^2} re^{-r^2/A^2}$, $\psi_{2p}^{(e)}(r) = \frac{4}{\sqrt{6}B^2} re^{-r/B}$ and $\psi_{2p}^{(t)}(r)$ to approximate $\psi_{2p}(r)$ (see Fig. 5(a)), and $\psi_{3d}^{(g)}(r) = \frac{2\sqrt{2}}{A^3} r^2 e^{-r^2/A^2}$, $\psi_{3d}^{(e)}(r) = \frac{4}{\sqrt{30}B^3} r^2 e^{-r/B}$ and $\psi_{3d}^{(t)}(r)$ to approximate $\psi_{3d}(r)$ (see Fig. 5(b)), which are used to fit the numerically calculated $r|\psi_{2p}(r)|^2$ and $r|\psi_{3d}(r)|^2$. Since the spatial extensions of 2p and 3d states are significantly larger than that of 1s state, the fittings to $r\left|\psi_{2p}^{(g)}(r)\right|^2$ and $r\left|\psi_{3d}^{(g)}(r)\right|^2$ are not as good as the

1s case generally. Nevertheless, $\psi_{2p}$ can still be well approximated by $\psi_{2p}^{(g)}$ under $d' \to \infty$, $r_0$ = 4.5 nm and $D \geqslant 6$ Å, and the approximation gets better for larger $D$ values. Under $d' = 5$ Å, $r_0 = 4.5$ nm, $\psi_{2p}$ is no longer well approximated by $\psi_{2p}^{(g)}$ but becomes closer to $\psi_{2p}^{(e)}$. On the other hand, for 3d state the exponential decay form $\psi_{3d}^{(e)}$ always agrees better with $\psi_{3d}$ than the gaussian form $\psi_{3d}^{(g)}$ for the shown range of $0 \leqslant D \leqslant 12$ Å. This is consistent with the results in Fig. 3(a-d), which indicate that the alignments of $E_{nm}$ and $\sqrt{\langle r^2 \rangle}$ for different exciton states have better agreement with the harmonic oscillator model under $d' \to \infty$ than those under $d' = 5$ Å. On the other hand, the transitional forms $\psi_{2p}^{(t)}$ and $\psi_{3d}^{(t)}$ again give nearly perfect fittings to $\psi_{2p}$ and $\psi_{3d}$, respectively. Thus, one can use $\psi_{2p}^{(t)}$ ($\psi_{3d}^{(t)}$) as a variational wave function to obtain the lowest-energy exciton state in the subspace of $m = \pm 1$ ($m = \pm 2$) with a high accuracy. We summarize our fitting results of $A$, $B$, $C$ and $C'$ to $\psi_{1s}$, $\psi_{2p}$ and $\psi_{3d}$ in Appendix II.

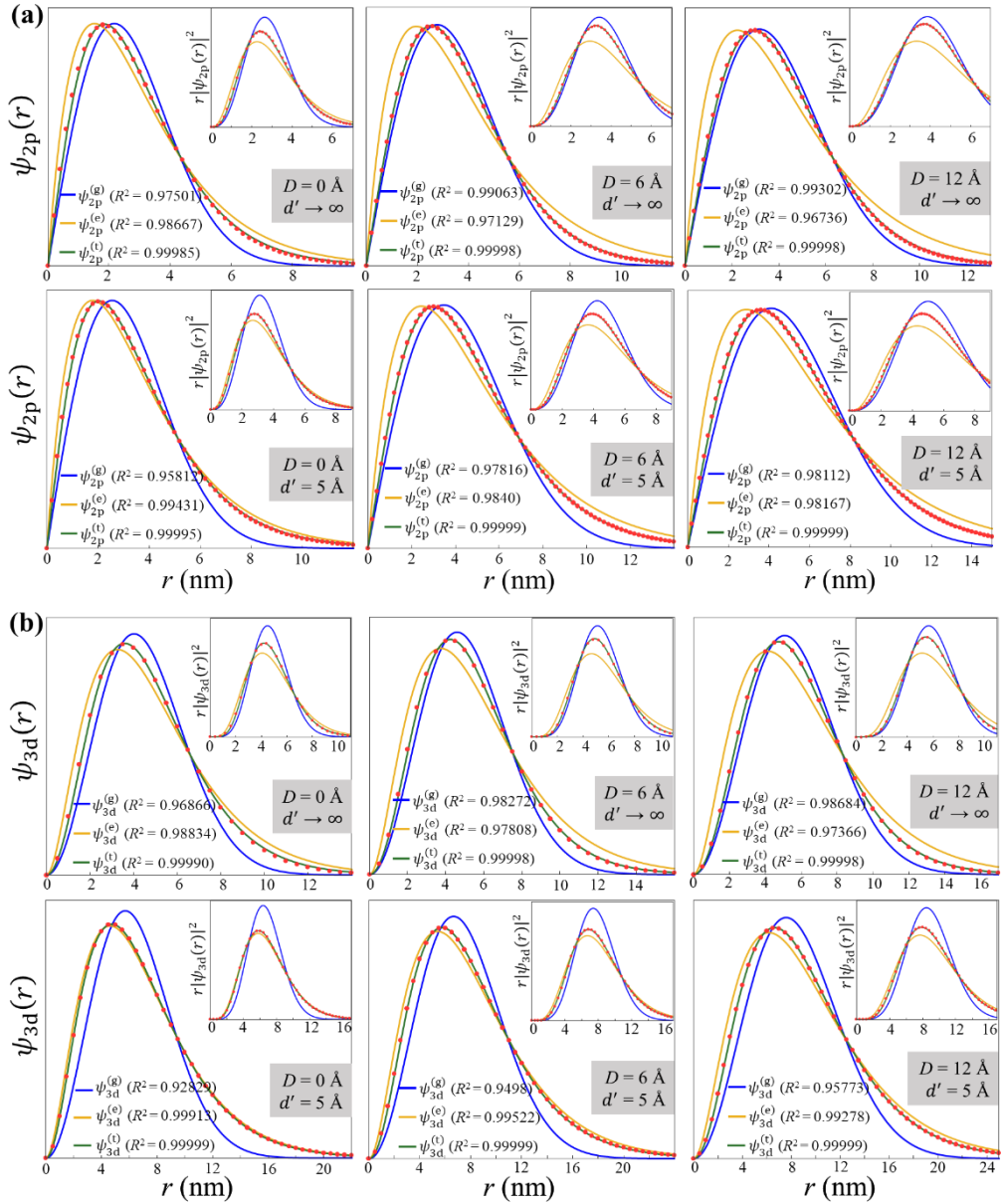

**Fig. 5** (a) The numerical data of $\psi_{2p}$ (red dots) and three approximated forms: $\psi_{2p}^{(g)}$ (blue lines), $\psi_{2p}^{(e)}$ (yellow lines) and $\psi_{2p}^{(t)}$ (green lines). The insets show fitting results to $r|\psi_{2p}(r)|^2$ (red dots) using $r\left|\psi_{2p}^{(g)}(r)\right|^2$ (blue lines), $r\left|\psi_{2p}^{(e)}(r)\right|^2$ (yellow lines) and $r\left|\psi_{2p}^{(t)}(r)\right|^2$ (green lines). (b) The numerical data of $\psi_{3d}$ (red dots) and three approximated forms: $\psi_{3d}^{(g)}$ (blue lines), $\psi_{3d}^{(e)}$ (yellow lines) and $\psi_{3d}^{(t)}$ (green lines). The insets show fittings to $r|\psi_{3d}(r)|^2$ (red dots) using $r\left|\psi_{3d}^{(g)}(r)\right|^2$ (blue lines), $r\left|\psi_{3d}^{(e)}(r)\right|^2$ (yellow lines) and $r\left|\psi_{3d}^{(t)}(r)\right|^2$ (green lines).

Finally, we consider inter-orbital couplings between 1s, 2s, 2p and 3d states which can be induced by an external electrostatic potential [34,35]. The coupling strengths are characterized by the inter-orbital transition dipole moments

$$\mathcal{D}_{1s \to 2p} \equiv \langle \Psi_{2,\pm 1} | \hat{r}_\pm | \Psi_{1,0} \rangle = \int_0^\infty dr r^2 \psi_{2p}^*(r) \psi_{1s}(r),$$

$$\mathcal{D}_{2p \to 2s} \equiv \langle \Psi_{2,0} | \hat{r}_\pm | \Psi_{2,\mp 1} \rangle = \int_0^\infty dr r^2 \psi_{2s}^*(r) \psi_{2p}(r), \qquad (6)$$

$$\mathcal{D}_{2p \to 3d} \equiv \langle \Psi_{3,\pm 2} | \hat{r}_\pm | \Psi_{2,\pm 1} \rangle = \int_0^\infty dr r^2 \psi_{3d}^*(r) \psi_{2p}(r).$$

Here $\hat{r}_\pm \equiv \hat{x} \pm i\hat{y}$. With the increase of $D$ and $r_0$, the spatial distribution of the wave function extends and all the above inter-orbital transition dipoles become larger (Fig. 6(a,b)). The strengths of these inter-orbital transition dipoles are in the order of several nm or several hundred Debye, one order of magnitude larger than the conduction-to-valence inter-band transition dipole in monolayer TMDs. Thus, an in-plane electric field with a strength of 10 V/μm can result in an ~ 30 meV coupling strength between 2s and 2p states, comparable to their energy separation (~ 10-30 meV for IXs in bilayer TMDs, see Fig. 3(a,c)). This can then result in a strong hybridization between 2s and 2p, as evidenced by recent experiments [34,45,46]. Such a behavior is also consistent with the observation that the 2s state can serve as a sensitive probe for external perturbations [12,43]. We note that in bilayer TMDs the 1s-2p energy separation for IXs is only ~ 50 meV, this implies that the IX ground state can become the strong hybridization of 1s and 2p under an in-plane electric field strength of several tens V/μm.

To summarize, we have shown that the large interlayer separation $D$ and screening length $r_0$ of bilayer TMDs result in the 1s wave function of IX well described by a gaussian form rather than the exponential decay form. In TMDs bilayers without hBN encapsulation, the energy of 2s state becomes more and more close to that of 3d state with the increase of interlayer separation, which satisfies $E_{2s} - E_{2p} \approx E_{3d} - E_{2p} \approx E_{2p} - E_{1s}$. For general parameters of $D$ and with the presence of hBN encapsulation where IX wave functions show neither gaussian nor exponential decay form, we give several simple analytic functions which smoothly cross from gaussian to exponential decay forms at different positions. These transitional forms are found to always fit nearly perfectly to ground and excited state wave functions, thus can serve as highly accurate variational functions. They can be used to obtain accurate estimations of various IX quantities, thus facilitate excitonic device applications.

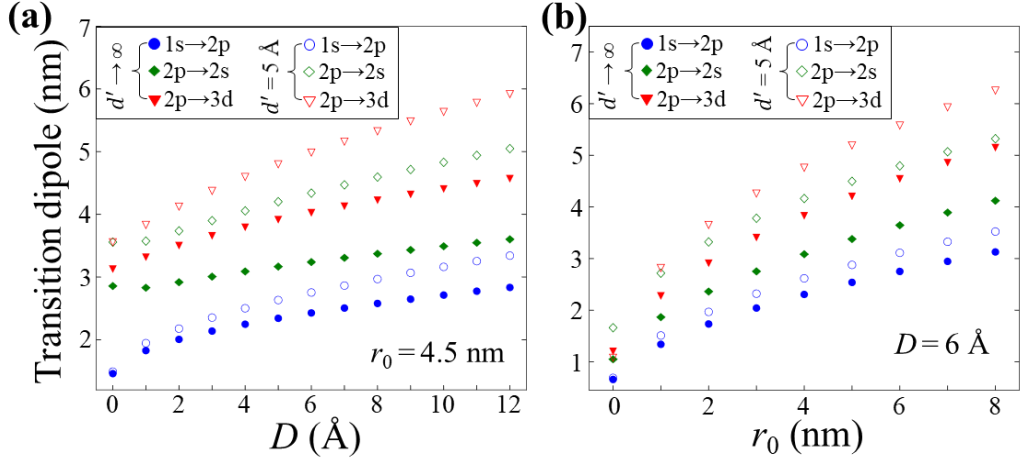

**Fig. 6** (a) The strengths of inter-orbital transition dipole moments $\mathcal{D}_{1s\to 2p}$, $\mathcal{D}_{2p\to 2s}$ and $\mathcal{D}_{2p\to 3d}$ as functions of $D$, under $r_0 = 4.5$ nm and $d' \to \infty$ or $d' = 5$ Å. (b) The inter-orbital transition dipole moments as functions of $r_0$ under $D = 6$ Å and $d' \to \infty$ or $d' = 5$ Å.

**Acknowledgements:** H.Y. acknowledges support by NSFC under grant No. 12274477, and the Department of Science and Technology of Guangdong Province in China (2019QN01X061).

## Appendix I. The interlayer Coulomb interaction under the dielectric screening of the substrate and capping-layer

As shown in Fig. 1(a), we set the vertical ($z$) positions of the two TMDs layers at $z_{1,2} = \pm D/2$. The substrate and the capping layer are infinitely thick whose dielectric constants are both given by $\epsilon$. The top (bottom) surface of the substrate (capping layer) is at $z_3 = -D/2 - d'$ ($z_4 = D/2 + d'$). Consider an electron located at $(\mathbf{r}, z) = (0, -D/2)$ in the lower TMD layer, the Poisson equation for the 3D electrostatic potential $\phi(\mathbf{r}, z)$ is

$$\delta(\mathbf{r})\delta\left(z + \frac{D}{2}\right) = \frac{1}{4\pi}\nabla_{\parallel}^2\phi(\mathbf{r}, z) + \frac{1}{4\pi}\frac{\partial^2\phi(\mathbf{r}, z)}{\partial z^2}$$
$$+ \kappa\nabla_{\parallel}^2\phi\left(\mathbf{r}, -\frac{D}{2}\right)\delta\left(z + \frac{D}{2}\right) + \kappa\nabla_{\parallel}^2\phi\left(\mathbf{r}, \frac{D}{2}\right)\delta\left(z - \frac{D}{2}\right).$$

Applying the 2D Fourier transform $\phi(\mathbf{r}, z) = \frac{1}{(2\pi)^2}\int d\mathbf{q}\, e^{i\mathbf{q}\cdot\mathbf{r}}\phi(\mathbf{q}, z)$, we get

$$\left(\frac{d^2}{dz^2} - q^2\right)\phi(\mathbf{q}, z) = 4\pi\delta\left(z + \frac{D}{2}\right) + 2r_0q^2\left[\phi\left(\mathbf{r}, -\frac{D}{2}\right)\delta\left(z + \frac{D}{2}\right) + \phi\left(\mathbf{r}, \frac{D}{2}\right)\delta\left(z - \frac{D}{2}\right)\right],$$

with $r_0 = 2\pi\kappa$ the screening length of the monolayer TMDs. The solution of $\phi(\mathbf{q}, z)$ can be determined from boundary conditions at interfaces located at $z_n$ ($n = 1,2,3,4$), given by $\phi(\mathbf{q}, z_n - 0^+) = \phi(\mathbf{q}, z_n + 0^+)$, and

$$\left.\frac{\partial\phi(\mathbf{q}, z)}{\partial z}\right|_{z=D/2+0^+} - \left.\frac{\partial\phi(\mathbf{q}, z)}{\partial z}\right|_{z=D/2-0^+} = 2r_0q^2\phi\left(\mathbf{q}, \frac{D}{2}\right),$$

$$\left.\frac{\partial\phi(\mathbf{q}, z)}{\partial z}\right|_{z=-D/2+0^+} - \left.\frac{\partial\phi(\mathbf{q}, z)}{\partial z}\right|_{z=-D/2-0^+} = 2r_0q^2\phi\left(\mathbf{q}, -\frac{D}{2}\right) - 4\pi,$$

$$\epsilon \frac{\partial \phi(\mathbf{q},z)}{\partial z}\bigg|_{z=D/2+d'+0^+} = \frac{\partial \phi(\mathbf{q},z)}{\partial z}\bigg|_{z=D/2+d'-0^+},$$

$$\epsilon \frac{\partial \phi(\mathbf{q},z)}{\partial z}\bigg|_{z=-D/2-d'-0^+} = \frac{\partial \phi(\mathbf{q},z)}{\partial z}\bigg|_{z=-D/2-d'+0^+}.$$

These boundary conditions require the electrostatic potential $\phi(\mathbf{q},z)$ for $z \in \left(-\frac{D}{2}-d', \frac{D}{2}+d'\right)$ to be in the form

$$\phi(\mathbf{q},z) = \left[-\frac{2\pi}{q} - qr_0\phi\left(\mathbf{q},-\frac{D}{2}\right)\right]e^{-q\left|z+\frac{D}{2}\right|} - qr_0\phi\left(\mathbf{q},\frac{D}{2}\right)e^{-q\left|z-\frac{D}{2}\right|} + \alpha(q)e^{qz} + \beta(q)e^{-qz}.$$

By setting $z = \pm D/2$ in the above equation we get

$$\alpha(q) = \frac{\left(\frac{\epsilon-1}{\epsilon+1}\right)^2 e^{-q(D/2+2d')}\left[\frac{2\pi}{q} - qr_0(1+e^{-qD})\phi\left(\mathbf{q},-\frac{D}{2}\right)\right]}{\left(\frac{\epsilon-1}{\epsilon+1}\right)^2 e^{-2q(D/2+d')} - e^{2q(D/2+d')}}$$

$$- \frac{\frac{\epsilon-1}{\epsilon+1}e^{\frac{qD}{2}}\left[\frac{2\pi}{q}e^{-qD} - qr_0(1+e^{-qD})\phi(\mathbf{q},-D/2)\right]}{\left(\frac{\epsilon-1}{\epsilon+1}\right)^2 e^{-2q(D/2+d')} - e^{2q(D/2+d')}}.$$

$$\beta(q) = \frac{\left(\frac{\epsilon-1}{\epsilon+1}\right)^2 e^{-q(D/2+2d')}\left[\frac{2\pi}{q}e^{-qD} - qr_0(1+e^{-qD})\phi\left(\mathbf{q},-\frac{D}{2}\right)\right]}{\left(\frac{\epsilon-1}{\epsilon+1}\right)^2 e^{-2q(D/2+d')} - e^{2q(D/2+d')}}$$

$$- \frac{\frac{\epsilon-1}{\epsilon+1}e^{qD/2}\left[\frac{2\pi}{q} - qr_0(1+e^{-qD})\phi\left(\mathbf{q},-\frac{D}{2}\right)\right]}{\left(\frac{\epsilon-1}{\epsilon+1}\right)^2 e^{-2q(D/2+d')} - e^{2q(D/2+d')}}.$$

The momentum-space form of the interlayer Coulomb potential $V(q) \equiv \phi(\mathbf{q}, D/2)$ is then

$$V(q) = \frac{-2\pi e^{-qD}MQ}{q[Q + (N - e^{-qD}M)qr_0][Q + (N + e^{-qD}M)qr_0]},$$

with

$$M = \left(1 - \frac{\epsilon-1}{\epsilon+1}e^{-2qd'}\right)^2,$$
$$Q = \left(1 - \frac{\epsilon-1}{\epsilon+1}e^{-q(D+2d')}\right)\left(1 + \frac{\epsilon-1}{\epsilon+1}e^{-q(D+2d')}\right),$$
$$N = \left(1 - \frac{\epsilon-1}{\epsilon+1}e^{-2q(D+d')}\right)\left(1 - \frac{\epsilon-1}{\epsilon+1}e^{-2qd'}\right).$$

The electron-hole Coulomb interaction takes the following real-space form:

$$V(r) = \frac{1}{2\pi}\int_0^\infty dq\, q J_0(qr) V(q).$$

Note that when the distance $d' \to \infty$, the Coulomb interaction will reduce to the case that the electron and hole are in a suspended bilayer, given by

$$V(r)|_{d'\to\infty} = -\int_0^\infty dq \frac{e^{-qD}J_0(qr)}{(1+r_0q)^2 - r_0^2 q^2 e^{-2qD}},$$

which is exactly the interlayer Coulomb potential obtained in Ref. [22] under $\epsilon = 1$.

# Appendix II. Fitting results for interlayer exciton wave functions

Table I. The fitting parameters for 1s states in Fig. 4 using Eq. (4) and 2p/3d states in Fig. 5 using Eq. (5). The units for *A*, *B*, *C* and *C'* are nm.

|  |  | $d' \to \infty$, $r_0$ = 4.5 nm | | | | $d'$ = 5 Å, $r_0$ = 4.5 nm | | | |
|---|---|---|---|---|---|---|---|---|---|
|  |  | $D$ = 0 | $D$ = 3 Å | $D$ = 6 Å | $D$ = 12 Å | $D$ = 0 | $D$ = 3 Å | $D$ = 6 Å | $D$ = 12 Å |
| 1s | *A* | 2.0047 | 2.9344 | 3.3393 | 3.9096 | 2.3831 | 3.0230 | 3.4417 | 4.1203 |
| | *B* | 1.4021 | 2.0654 | 2.3514 | 2.7543 | 1.6641 | 2.1233 | 2.4199 | 2.8999 |
| | *C* | 1.0226 | 2.9162 | 3.6956 | 4.8813 | 0.9393 | 1.8715 | 2.4253 | 3.3609 |
| | *C'* | 1.0810 | 1.1496 | 1.2203 | 1.3144 | 1.3876 | 1.5168 | 1.6297 | 1.8116 |
| 2p | *A* | 3.0730 | 3.5805 | 3.9434 | 4.4804 | 4.4244 | 4.6750 | 4.9810 | 5.5652 |
| | *B* | 1.5237 | 1.7811 | 1.9626 | 2.2309 | 2.1883 | 2.3164 | 2.4710 | 2.7644 |
| | *C* | 2.0727 | 3.5535 | 4.4015 | 5.6265 | 1.8192 | 2.2990 | 2.8141 | 3.7249 |
| | *C'* | 1.2326 | 1.2331 | 1.2790 | 1.3596 | 1.9891 | 2.0362 | 2.1027 | 2.2407 |
| 3d | *A* | 4.0273 | 4.2958 | 4.5953 | 5.0875 | 6.9950 | 7.1160 | 7.2894 | 7.7019 |
| | *B* | 1.6335 | 1.7455 | 1.8682 | 2.0693 | 2.8290 | 2.8791 | 2.9511 | 3.1214 |
| | *C* | 3.1084 | 4.1320 | 5.0004 | 6.2782 | 2.5494 | 2.7932 | 3.1745 | 4.0121 |
| | *C'* | 1.3603 | 1.3500 | 1.3730 | 1.4337 | 2.6913 | 2.7191 | 2.7542 | 2.8388 |